# Les réservations et les suspensions de cotation sont-elles un frein à l'efficience informationnelle des marchés ?

Karine Michalon

Le lundi 19 octobre 1987, à Wall Street, en une seule journée, l'indice Dow Jones, en passant de 2 246 points à 1 738 a perdu 508 points. Il recule ainsi de 22,6% soit presque le double de ce qu'avait engendré la crise de 1929. Ce krach a insufflé la volonté de mettre en place des mécanismes d'interruptions de cotation capables de corriger l'incapacité apparente du marché à traiter des quantités importantes d'informations dans des délais très courts. Aujourd'hui, les réglementations de toutes les places boursières prévoient des procédures d'interruption de cotation. Selon le marché et la réglementation en vigueur, ces ruptures peuvent concernées l'ensemble des transactions ou uniquement un titre. Les mesures prises au niveau d'un seul titre sont incontestablement le type de réglementation le plus ancien et le plus universel. Elles peuvent prendre soit la forme de réservations soit celle de suspensions. Les *réservations* ont lieu sur un titre quand son cours approche une certaine limite appelée seuil de réservation. Elles sont appliquées sur de nombreux marchés particulièrement les marchés émergents. Les modalités d'application, les seuils et durée de réservation varient d'une place à l'autre. Les *suspensions* sont exceptionnelles et concomitantes à des événements spéciaux surgissant dans la vie de la société émettrice. Elles représentent la forme la plus répandue des interruptions individuelles de cotation. Elles sont employées sur toutes les places boursières. La plupart sont motivées par l'attente ou l'annonce d'une information, d'autres par un déséquilibre entre l'offre et la demande, la conformité avec les textes de lois, des irrégularités ou encore une insuffisance de titres à la disposition du public ou un niveau d'activité insuffisant. Elles peuvent intervenir à la demande des régulateurs de marché, de la société ou de la bourse elle-même.

Sur tous les marchés, le rôle annoncé de ces interruptions, est de garantir l'efficience informationnelle des marchés. Il s'agit de permettre l'information des participants, réduire la volatilité, garantir la transparence et la liquidité des marchés. Mais le bien-fondé de ces mécanismes a très vite suscité la polémique. Les détracteurs des suspensions pensent que ce sont des barrières à l'échange. Lors de la réouverture, l'intérêt des investisseurs n'est pas forcément de révéler immédiatement l'information qu'ils détiennent, ce qui peut conduire à un bruitage temporaire des prix. Par ailleurs, l'accroissement de la couverture médiatique de l'interruption amène de nouveaux investisseurs et peut être un facteur d'éclatement de consensus, de dispersion accrue des anticipations et donc d'accroissement de la volatilité. Les bienfaits des réservations seraient de diminuer la volatilité en imposant un plancher et un plafond aux prix au sein d'une même séance et de fournir une période d'accalmie sur les marchés. Les désavantages des réservations seraient d'accroître la volatilité et le volume et de perturber la liquidité. Un autre inconvénient serait d'induire un « effet gravitationnel ». Dès lors que les cours se rapprochent du seuil de déclenchement de la réservation, le mécanisme peut inciter les cours à atteindre plus rapidement la limite, étant donné que les opérateurs dénouent précipitamment leurs positions avant la fermeture du marché afin d'éliminer l'incertitude entourant la valeur de leurs actifs.

Mais qu'en est-il en pratique ? Les ruptures de cotation sont-elles efficaces ? Aucune étude empirique ne permet de l'affirmer. Citons quelques exemples. Kabir (1994)[1] montre, à la Bourse de Londres, qu'il n'y a pas une dissémination de l'information à travers tous les investisseurs pendant les suspensions. Lee et al (1994)[2] montrent que le volume et la volatilité suivant les suspensions du NYSE sont significativement plus importants dans la journée d'échanges suivant la halte. Ces résultats sont confirmés par Corwin et Lipson (2000)[3]. L'efficacité des réservations n'est pas non plus prouvée. Chan et al (2005)[4] étudient les réservations à la bourse de Kuala Lumpur en Malaisie. Les auteurs montrent que les limites de prix n'améliorent pas l'asymétrie d'information et retardent l'arrivée d'information. Ils prouvent qu'un effet gravitationnel exacerbe le déséquilibre dans les ordres. Kim et Rhee (1997)[5] montrent que les réservations à la Bourse de Tokyo ne sont pas efficaces pour contenir des effets de sur réaction et diminuer la volatilité. De même Chen (1993)[6] à Taiwan, Phylaktis et al (1999)[7] en Grèce, Bildik et Gülay (2006)[8] en Turquie ne montrent pas que les limites de prix réduisent la volatilité. Ces derniers montrent également une interférence dans les échanges. Réservations et suspensions semblent donc être aujourd'hui, sur la plupart des places boursières, des freins à l'efficience informationnelle des marchés. Les autorités réglementaires ont-elles intérêt à maintenir ces mécanismes ?

---

[1] Kabir R., 1994, Share price behaviour around trading suspensions on the London Stock Exchange, Applied Financial Economics, 24, 289-295.

---

[2] Lee, Ready et Seguin, 1994, Volume, Volatility, and New York Stock Exchange Trading Halts, The journal of Finance, 49(1), 183-214.

[3] Corwin S., et M. Lipson, 2000, Order flow and liquidity around NYSE trading halts, Journal of Finance, 55, 1771 – 1801.

[4] Chan S. H., K. A. Kim, and S. G. Rhee, 2005, Price limit performance: evidence from transactions data and the limit order book, Journal of empirical Finance 12, 269-290.

[5] Kim K.A. et S.G. Rhee, 1997, Price limits performance: evidence from Tokyo Stock Exchange, Journal of Finance, 52, 885-901.

[6] Chen, Y.-M, 1997, Price limits and liquidity: a five-minute data analysis, Journal of Financial Studies 4, 45-65.

[7] Phylaktis, K., Kavussanos, M., Manalis, G., 1999, Price limits and stock market volatility in the Athens Stock Exchange, European Financial Management 5, 69-84.

[8] Bildik, R., and Gülay, G. 2006, Are price limits effective? Evidence from the Istanbul Stock Exchange, Journal of Financial Research, 29(3), 383-403.